\documentclass[12pt,reqno]{article}
\usepackage{graphicx}
\usepackage{amsmath}
\usepackage{times}
\usepackage{hyperref}
\usepackage{float}
\usepackage{accsupp}
\usepackage{fullpage}
\usepackage{subcaption} 
\usepackage{caption}
\usepackage{longfigure}

\begin{document}

\title{
Optimization of Retrieval-Augmented Generation Context with Outlier Detection 
\\\
\\\
\small NextAI Systems LLC \normalsize
}
\author{
Vitaly Bulgakov \href{mailto:vbulgakov@nextaisystems.com}
{\BeginAccSupp{method=escape,ActualText={}}vbulgakov@nextaisystems.com\EndAccSupp{}}
}

\date{\today}

\maketitle

\begin{abstract}
In this paper, we focus on methods to reduce the size and improve the quality of the prompt context required for question-answering systems. Attempts to increase the number of retrieved chunked documents and thereby enlarge the context related to the query can significantly complicate the processing and decrease the performance of a Large Language Model (LLM) when generating responses to queries. 
It is well known that a large set of documents retrieved from a database in response to a query may contain irrelevant information, which often leads to hallucinations in the resulting answers. Our goal is to select the most semantically relevant documents, treating the discarded ones as outliers. We propose and evaluate several methods for identifying outliers by creating features that utilize the distances of embedding vectors, retrieved from the vector database, to both the centroid and the query vectors.
The methods were evaluated by comparing the similarities of the retrieved LLM responses to ground-truth answers obtained using the OpenAI GPT-4o model. It was found that the greatest improvements were achieved with increasing complexity of the questions and answers.

\end{abstract}

\subsection*{1. Introduction}

Improving context retrieval in question-answering systems has garnered considerable attention over the past few years. For instance, \cite{1} presents a comprehensive survey on the Retrieval-Augmented Generation (RAG) framework, detailing the involved retrievers and generators. It explores various enhancement methods for RAG, evaluates benchmark frameworks, and addresses current limitations and future directions, providing a solid foundation for understanding RAG techniques and their applications.
In another study, \cite{2} introduces xRAG, a method for extreme context compression that reinterprets document embeddings used in dense retrieval as features for language models. This approach aims to maintain the semantic coherence of retrieved documents while minimizing memory overhead and computational expense. xRAG employs techniques such as self-distillation and instruction tuning to enhance the use of contextual information.
Also, \cite{3} presents a novel in-context retrieval approach for RAG systems that avoids the traditional chunking process. It uses encoded hidden states of documents for retrieval, improving the fidelity and accuracy of the evidence text used for generating responses. This method ensures that relevant and precise context is maintained without disrupting the semantic coherence of the documents.
\cite{4} discusses the concept of re-ranking, where a re-ranking agent re-orders retrieved documents based on additional factors such as user behavior, document popularity, or deeper semantic analysis, ensuring the most useful and relevant documents appear at the top of search results. In \cite{5} we explore the benefits of reducing vector database dimensions with Fast Fourier Transform, with a focus on computational efficiency in processing context documents. 
\\\
In the described approach, we propose a mechanism to improve the quality and reduce the size of the context based on retrieved documents for a given query. This mechanism identifies outliers that are far from or irrelevant to the query and focuses on the portion of the context most related to the query. Distances are calculated for each retrieved document, represented as an embedding vector by a sentence transformer model. We utilize distances to the vectors’ centroid and the query vector. These distances form features based on the following methods: \textit{concatenation, weighted sum, interaction, and polynomial}, which will be described later. The constructed features provide a rich representation of the distances between embedding vectors, the centroid, and the query vector, capturing various aspects of the relationship between the vectors and helping distinguish typical vectors from outliers. Outliers are determined with help of a Gaussian Mixture Model (GMM) and Log-Likelihood approach with selected percentile, \cite{6} and \cite{7}. 
\\\
We experimented with three categories of questions of increasing complexity and demonstrated that the approach provides the most significant advantage with the most complex questions.

\subsection*{2. Theory and Methods}

Our approach aims to identify outliers among a set of embedding vectors used in a context for Retrieval-Augmented Generation (RAG). The outliers are detected based on distances from a query vector and a centroid vector. Below is a detailed explanation of the methods implemented, along with the corresponding formulas.

\paragraph{Distance Calculation}
For each embedding vector in the set, the distances to a given query vector and a centroid vector are calculated:
\begin{equation}
    \textbf{Distance to Centroid}:
    d_{\text{centroid}} = \| \mathbf{v}_i - \mathbf{c} \|
\end{equation}
\begin{equation}
    \textbf{Distance to Query Vector}:
    d_{\text{query}} = \| \mathbf{v}_i - \mathbf{q} \|
\end{equation}

Where:
\begin{itemize}
    \item \(\mathbf{v}_i\) is the \(i\)-th embedding vector,
    \item \(\mathbf{c}\) is the centroid vector,
    \item \(\mathbf{q}\) is the query vector,
    \item \(\|\cdot\|\) denotes the Euclidean norm.
\end{itemize}

We use a weighting factor \(0 \le \alpha \le 1\) to balance these distances:
\begin{equation}
\text{distance to centroid (weighted)} = d_{\text{centroid}} \times (1 - \alpha)
\end{equation}
\begin{equation}
\text{distance to query (weighted)} = d_{\text{query}} \times \alpha
\end{equation}

\paragraph{Feature Creation Methods}

We used multiple methods to combine these distances into features (see, e.g. \cite{8} - \cite{9}) for further analysis:

\begin{equation}
\textbf{Concatenate}:
    \mathbf{f}_i = \left[ d_{\text{centroid}}, d_{\text{query}} \right]
\end{equation}
\begin{equation}
    \textbf{Weighted Sum}:
    \mathbf{f}_i = \alpha d_{\text{query}} + (1 - \alpha) d_{\text{centroid}}
\end{equation}
\begin{equation}
    \textbf{Interaction}:
    \mathbf{f}_i = \left[ d_{\text{centroid}}, d_{\text{query}}, d_{\text{centroid}} \times d_{\text{query}}, \frac{d_{\text{centroid}}}{d_{\text{query}} + \epsilon} \right]
\end{equation}
Where \(epsilon\) is a small value to avoid division by zero.
\begin{equation}
   \textbf{Polynomial}:
    \mathbf{f}_i = \text{PolynomialFeatures}(d_{\text{centroid}}, d_{\text{query}})
\end{equation}
This expands the distance features to include polynomial combinations up to the specified degree.

\paragraph{Standardization}

Standardization is applied to the feature vectors to ensure they have zero mean and unit variance:

\begin{equation}
\mathbf{f}_{\text{normalized}} = \frac{\mathbf{f}_i - \mu}{\sigma}
\end{equation}

Where \(\mu\) is the mean and \(\sigma\) is the standard deviation of the features.

\paragraph{Dimensionality Reduction}

If the number of features is more than two, we apply Principal Component Analysis (PCA) to reduce the features to two dimensions for visualization. We also use various PCA dimensions for outliers detection which will be explained further:

\begin{equation}
\mathbf{f}_{\text{PCA}} = \text{PCA}(\mathbf{f}_{\text{normalized}})
\end{equation}

\paragraph{Outlier Detection}

Outliers are identified based on a log-likelihood threshold. Points with a log-likelihood below the threshold are considered outliers.

\begin{enumerate}
    \item \textbf{Fit a Gaussian Mixture Model (GMM)}: 
    The feature vectors are modeled using a GMM, which assumes that the data is generated from a mixture of several Gaussian distributions. The GMM is parameterized by the mean vectors \(\mu_k\), covariance matrices \(\Sigma_k\), and mixture weights \(\pi_k\) for each component \(k\).

    \item \textbf{Calculate Log-Likelihood}: 
    For each feature vector \(\mathbf{f}_i\), compute the log-likelihood under the GMM:
    \begin{equation}
    \log p(\mathbf{f}_i) = \log \left( \sum_{k=1}^K \pi_k \mathcal{N}(\mathbf{f}_i | \mu_k, \Sigma_k) \right)
    \end{equation}
    where \(\mathcal{N}(\mathbf{f}_i | \mu_k, \Sigma_k)\) is the probability density function of the Gaussian distribution with mean \(\mu_k\) and covariance \(\Sigma_k\).

    \item \textbf{Determine Outliers}:
    Identify the outliers as those points whose log-likelihood is below a certain threshold. This threshold can be defined as a certain percentile of the log-likelihood values. For example, consider the bottom 10
    \begin{equation}
    \text{threshold} = \text{percentile}(\{\log p(\mathbf{f}_i)\}_{i=1}^N, 10)
    \end{equation}
    where \(N\) is the total number of feature vectors.

    \item \textbf{Outlier Identification}:
    Points with log-likelihood values less than the threshold are marked as outliers:
    \begin{equation}
    \text{outliers} = \{ \mathbf{f}_i \mid \log p(\mathbf{f}_i) < \text{threshold} \}
    \end{equation}
    GMM provides a probability distribution over clusters for each data point. This means each point can belong to multiple clusters with different probabilities. The choice of \(K\) (number of components) is crucial. It can be selected based on prior knowledge, or determined using model selection criteria like the Bayesian Information Criterion (BIC) or the Akaike Information Criterion (AIC), which balance model fit and complexity. Another approach would be by varying a combinations of GMM components and PCA dimensions one can control and optimize outliers selection. 
\end{enumerate}

\subsection*{3. Numerical study}

The purpose of the numerical study was to determine if better responses could be obtained for a query using a filtered context (excluding outlier documents) compared to using the context based solely on a set of documents retrieved from the retrieval language model. Our base text-to-text model was 'TinyLlama/TinyLlama-1.1B-Chat-v1.0,' which, although rather small, proved to be a good choice for research and allowed us to conduct extensive experiments on a regular laptop. We also tested the proposed method with the 'mistralai/Mistral-7B-Instruct-v0.2' model and obtained very similar results but at a much higher cost.
\\\
As a document retriever, we used a popular sentence transformer model, ‘sentence-transformers/all-mpnet-base-v2,’ which can be used as a text to dense vectors converter. With this model we converted all chunked documents and stored them in a FAISS vector database developed by Facebook Research \cite{10}. All models we used are publicly available in Hugging Face hub \cite{11}.
\\\
For testing, we used various datasets. The results of this study are obtained with SQuAD2.0 - The Stanford Question Answering Dataset, which includes a mixture of 35 topics covering political aspects, geography, mathematics, chemistry, religion, etc., ranging from “Amazon Rainforest” to “Yuan Dynasty,” from “Pharmacy” to “Prime Numbers.” To generate questions and corresponding answers for a benchmark, we leveraged the extensive knowledge and capabilities of the OpenAI GPT-4o model with the uploaded dataset text. The following three categories of questions and corresponding answers were generated:

\subsubsection*{“Simple”}
Questions that allow for a short simple answer, such as:
\begin{itemize}
    \item When did the 1973 oil crisis begin?
    \item What did Nixon request from Congress on October 19, 1973?
\end{itemize}

\subsubsection*{“Broader”}
Questions that require to provide detail and insight, such as:
\begin{itemize}
    \item What allowed the tropical rainforest to spread out across the continent after the Cretaceous–Paleogene extinction event?
    \item How did the construction of highways in the Amazon lead to deforestation?
\end{itemize}

\subsubsection*{“Double”}
Two questions of “broader” type combined in one, such as:
\begin{itemize}
    \item What caused the recurrence of plague outbreaks in Europe until the 19th century and how did the rise of the Andes Mountains create the Solimões Basin?
    \item What led to the creation of the National Energy Act of 1978 and what caused the Amazon rainforest to be considered unsustainable by 2100?
\end{itemize}

When splitting text documents into chunks we tried to select a max chunk size so that the number of tokens from 20 – 25 chunks, considered as documents, does not exceed the sequence length of the model, which in case of TinyLlama-1.1B-Chat-v1.0 model is 2048.

In Outlier Detection section we described the method of selecting outliers from a given set of feature vectors.  This selection depends on the number of GMM components (clusters). For example in case of "interaction" method the size of the feature vector \(\mathbf{f}_i\) is 4. This is because the feature vector is composed of four components:
\begin{enumerate}
    \item \(d_{\text{centroid}}\)
    \item \(d_{\text{query}}\)
    \item \(d_{\text{centroid}} \times d_{\text{query}}\)
    \item \(\frac{d_{\text{centroid}}}{d_{\text{query}} + \epsilon}\)
\end{enumerate}
Each of these components represents a different feature derived from the distances \(d_{\text{centroid}}\) and \(d_{\text{query}}\). Then, by projecting these 4 features to principal components using PCA and selecting 2, 3, or 4 components, another leverage for controlling the clustering mechanism is introduced. We will use the following notation in selecting outliers:  
\begin{itemize}
\item Number of clusters: [4, 5, 6]
\item PCA dimensions: [2, 3]
\end{itemize}
By running the double loop over the number of clusters and PCA dimensions we will be getting different sets of outlier documents and some of them will be common. The final set of outliers will be those that are common or those that that are common at least at certain given number of occurrences. We will call it "min outlier frequency" denoted by "freq".
\\\
We will use the following experiment setup:

\begin{itemize}
    \item \textbf{Text-to-text model:} 'TinyLlama/TinyLlama-1.1B-Chat-v1.0'
    \begin{itemize}
        \item \textbf{Max Sequence Length:} 2048
    \end{itemize}
    \item \textbf{Retrieval Model:} ‘sentence-transformers/all-mpnet-base-v2’
    \begin{itemize}
        \item \textbf{max\_seq\_length:} 384
        \item \textbf{sentence\_embedding\_dimension:} 768
    \end{itemize}
    \item \textbf{max\_new\_tokens:} 1500
    \item \textbf{Number of Docs:} 20
    \item \textbf{Threshold Percentile:} 15\%
    \item \textbf{Number of clusters:} [4, 5, 6]
    \item \textbf{PCA dimensions:} [2, 3]
\end{itemize}

We use the following simple template to form a query prompt to the model that includes a short instruction, context and the query itself:

\begin{verbatim}
formatted_prompt = f'''
    You are a friendly chatbot who responds to the user's question by 
    looking into context.</s>
    Context: 
    {context}
    </s>
    Question: {question}</s>
    '''
\end{verbatim}
Context consists of documents retrieved on a query from the vector database. We call it a "filtered prompt" when all outliers are removed from the context. The response to a filtered prompt is compared with the ground-truth response. The same is done with the "original prompt," where the context consists of the same number of first documents retrieved from the database. For comparison, we use two types of similarity metrics: cosine (dense vector) similarity and TFIDF similarity based on tokens. Our purpose is to determine the improvement received when using "filtered prompts" over "original prompts" with respect to these similarities. For query \( Q_i \), let \( \text{Rg}_i \) be the ground truth response, \( \text{Rf}_i \) be the filtered response, and \( \text{Ro}_i \) be the original response. The improvement in similarity can be expressed as the difference in similarities of \( \text{Rf}_i \) to \( \text{Rg}_i \) and \( \text{Ro}_i \) to \( \text{Rg}_i \) divided by the similarity of \( \text{Ro}_i \) to \( \text{Rg}_i \):

\begin{equation}
    \text{Improvement}_i = \frac{\text{Similarity}(\text{Rf}_i, \text{Rg}_i) - \text{Similarity}(\text{Ro}_i, \text{Rg}_i)}{\text{Similarity}(\text{Ro}_i, \text{Rg}_i)}
\end{equation}

Where:
\begin{itemize}
    \item \( \text{Similarity}(\text{Rf}_i, \text{Rg}_i) \) is the similarity between the filtered response and the ground truth response.
    \item \( \text{Similarity}(\text{Ro}_i, \text{Rg}_i) \) is the similarity between the original response and the ground truth response.
\end{itemize}
The average improvement over \(N\) questions can be calculated as follows:

\begin{equation}
    \text{Average Improvement} = \frac{1}{N} \sum_{i=1}^{N} \text{Improvement}_i
\end{equation}

Where:
\begin{itemize}
    \item \( N \) is the total number of questions.
    \item \( \text{Improvement}_i \) is the improvement for question \( Q_i \).
\end{itemize}

Table 1 on Figure 1 demonstrates the summary of conducted experiments. Columns "Emb \%" and "TFIDF \%" express positive or negative effect in similarities obtained in corresponding experiments or, in other words, average improvement given by (14)-(15) in the embedding vector (cosine) and textual TFIDF similarities respectively. The average values of these parameters are depicted as graphs as the number of processed questions increases in the appendix.  

\begin{figure}[H]
  \includegraphics[width=1.0\textwidth]{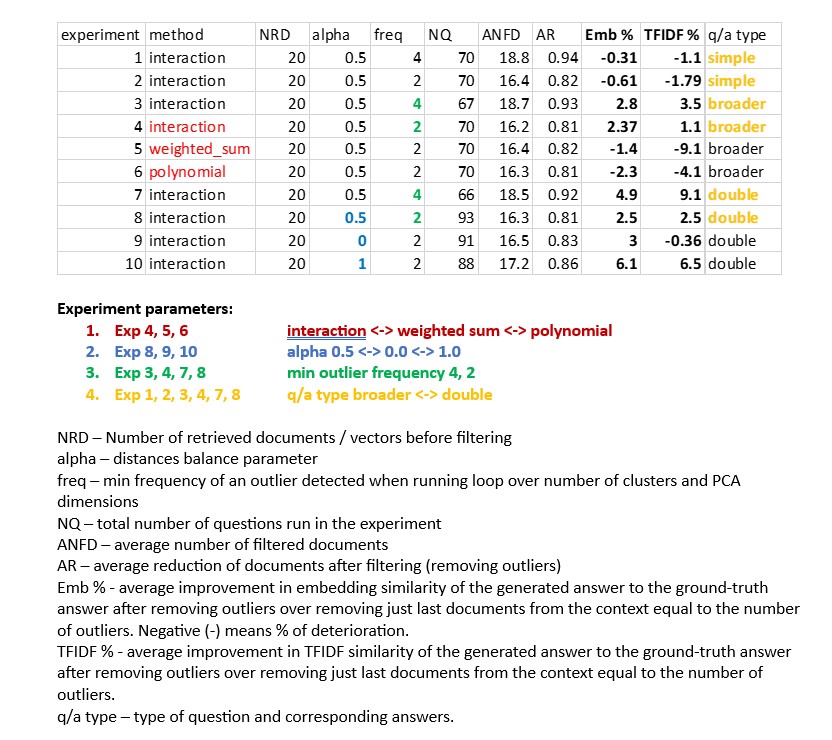}
  \caption{Summary of conducted experiments}
  \label{table:1}
\end{figure}

For illustrative purposes we generated images of clusters and outliers with 2 principal components of a feature vector, see (7), as coordinates. Figure 2 demonstrates clusters and outliers distribution for the "interaction" method (7) that has 4 features and with min\_outlier\_freq = 2.   

\begin{figure}[H]
  \includegraphics[width=0.8\textwidth]{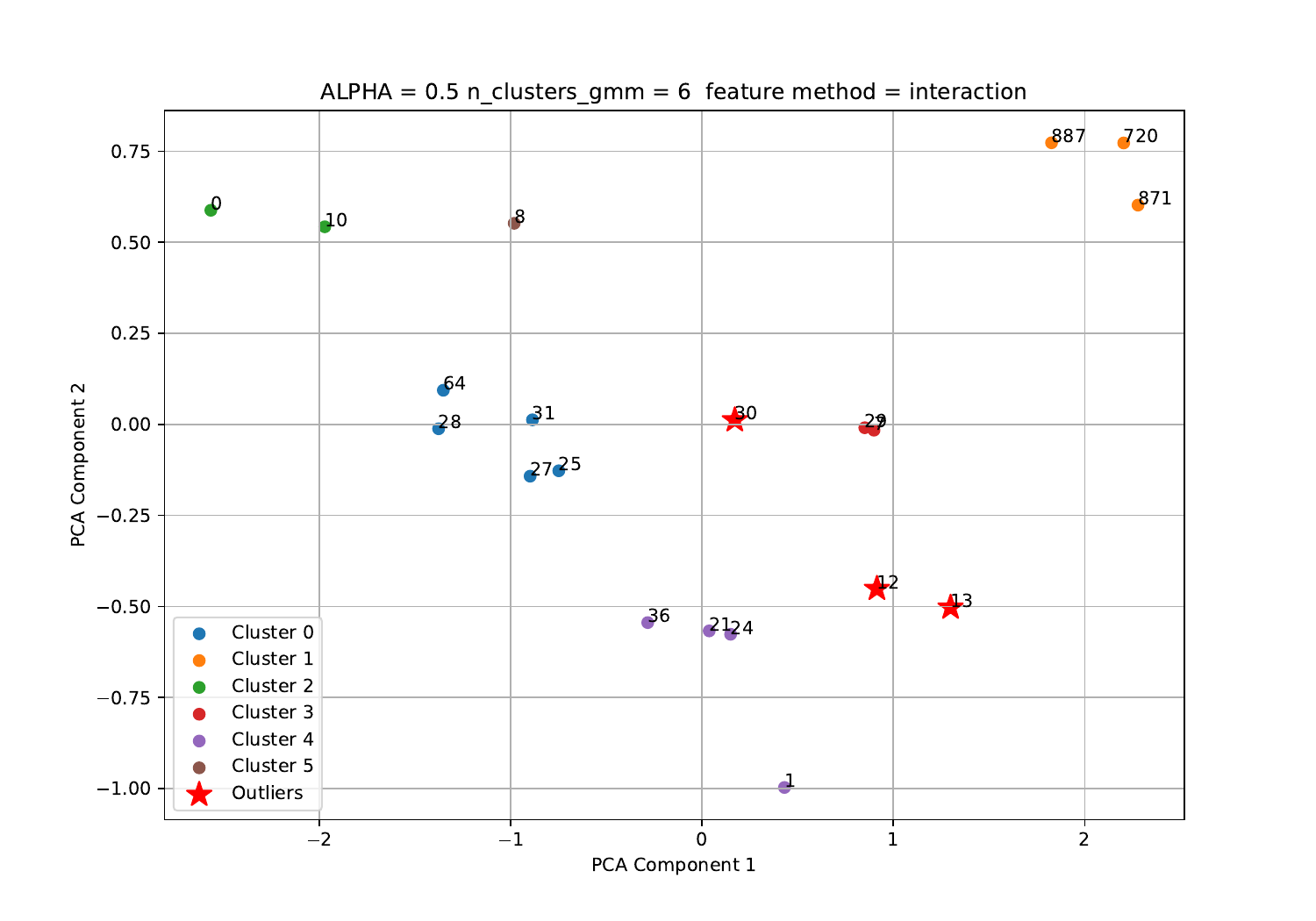}
  \caption{Clusters and outliers with 2 principal components of the  feature vector}
  \label{fig:2}
\end{figure}

The following 4 observations can be made from the results obtained:

\subsubsection*{Observation 1}
Experiment 4-6 show that the "interaction" method has an obvious advantage in term of both, dense vector Cosine and TFIDF similarities improvement, over "weighted sum" and "polynomial" methods. The rest of experiments were conducted with "interaction" method. A possible explanation could be that it has a largest number of features than other methods that could better represent clusters and outliers. 

\subsubsection*{Observation 2}
Experiments 3, 4 and 7, 8 demonstrate that the similarity improvement with a higher minimum outlier frequency of 4 is better than that with a lower frequency of 2. This is explained by a stronger outliers filtering. But the flip side of this improvement is a lower reduction of the number of filtered documents which causes higher computational cost in model's response retrieval than that with the lower frequency. 

\subsubsection*{Observation 3}
In (3)-(4) we use parameter \(\alpha\) that balances contribution of 2 types of distances (1)-(2) into features. The higher the \(\alpha\) the more significant role is played by the distance to query. 
Experiments 8-10 demonstrated that the higher \(\alpha\) is advantageous, at least in this particular testing settings.

\subsubsection*{Observation 4}
This is probably the most important observation. As was said, we introduced 3 categories of questions and corresponding answers. Let's take a look at Figure 3 that demonstrates how the similarity improvement depends on the questions complexity:
\begin{figure}[H]
  \includegraphics[width=0.8\textwidth]{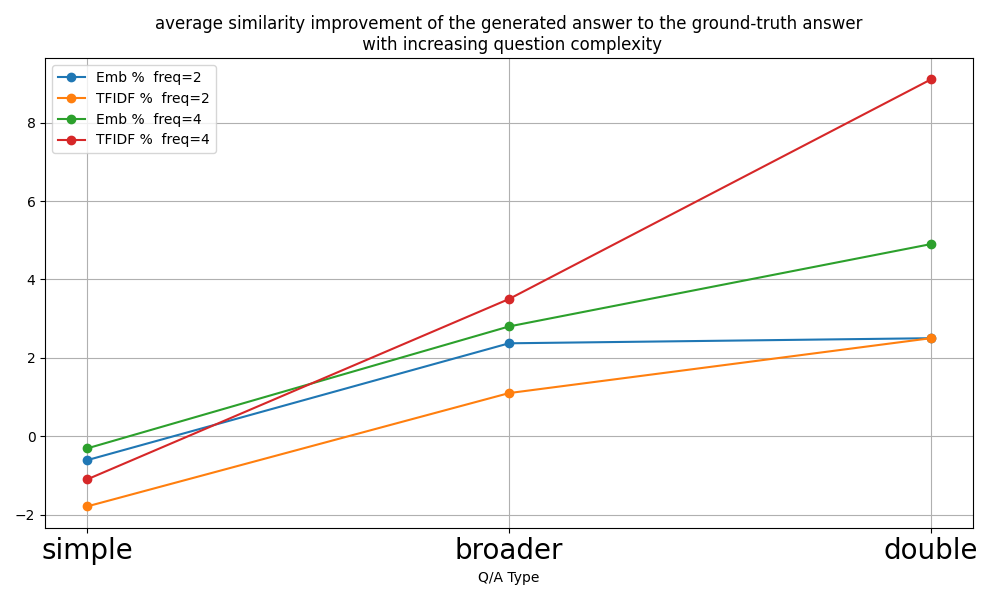}
  \caption{Similarity improvement with increasing question complexity}
  \label{fig:3}
\end{figure}

\textbf{The best improvement is achieved with the most complex question structure}. The explanation of this phenomena could be the following:
\\\
Simple, short queries generally do not require significant improvement, resorting, or filtering of retrieved documents for several reasons:

1. \textbf{Clarity and Specificity}:
   Simple queries are usually clear and specific, making it easier for retrieval models to find relevant documents. There is less ambiguity in understanding what the user is looking for, leading to more accurate initial retrieval.

2. \textbf{Limited Scope}:
   Short queries typically have a limited scope, focusing on a single piece of information or a straightforward fact. This reduces the chance of retrieving a large number of irrelevant documents, as the search space is more confined.

3. \textbf{Reduced Ambiguity}:
   Simple queries often contain fewer ambiguous terms. For instance, a query like "capital of France" is very specific compared to a broader or more complex query like "effects of climate change on urban development." Reduced ambiguity helps in retrieving precise documents without needing additional filtering or resorting.

4. \textbf{High Precision}:
   Retrieval models are generally well-optimized for handling straightforward, fact-based queries. The top results for such queries are usually highly relevant, and additional processing may not significantly improve the relevance of the results.

5. \textbf{Efficiency}:
   For short, simple queries, the computational cost of additional processing (such as filtering outliers or resorting documents) may not be justified. The initial retrieval is often sufficient to provide the necessary information, making further refinement unnecessary.

For example a simple query like "What is the birthdate of Albert Einstein." This query is:
\begin{itemize}
    \item \textbf{Clear and specific}: It asks for a specific piece of information.
    \item \textbf{Limited in scope}: It only requires one fact, which reduces the search space.
    \item \textbf{Unambiguous}: The terms used are clear and not open to multiple interpretations.
\end{itemize}

In contrast, \textbf{a more complex query like "impact of climate change on coastal cities and mitigation strategies" involves multiple concepts and relationships, leading to a broader and more diverse set of documents. This complexity increases the likelihood of retrieving less relevant documents, making resorting and filtering more beneficial}.

\subsection*{4. Conclusion}

In conclusion, this study demonstrates the efficacy of filtering outlier documents to improve the quality of responses in question-answering systems using retrieval-augmented generation (RAG). By removing documents that are semantically irrelevant or far from the query, we have shown that the context can be significantly refined, leading to more accurate and relevant answers. Besides, a reduced context length is beneficial it terms of computational cost required to produce the response by the model. 

Our experiments with different text-to-text models, including 'TinyLlama/TinyLlama-1.1B-Chat-v1.0' and 'mistralai/Mistral-7B-Instruct-v0.2,' revealed that even smaller models benefit from this approach, making it feasible to conduct extensive research on standard hardware. The use of a sentence transformer model, 'sentence-transformers/all-mpnet-base-v2,' for document retrieval further reinforced the robustness of our method across diverse datasets, including the SQuAD2.0 dataset.

The key observations from our experiments indicated that:
1. The "interaction" method provided the most significant improvement in similarity metrics.
2. Higher "min outlier frequency" led to better similarity improvements, albeit at the cost of increased computational demands.
3. The balance parameter \(\alpha\) played a certain role in enhancing the feature representation for outlier detection.
4. The complexity of questions had a profound impact on the effectiveness of the filtering mechanism, with more complex queries benefiting the most from the refined context.

Overall, this study highlights the potential for enhanced context retrieval methods to address challenges in RAG systems, particularly when dealing with complex queries. Future work will focus on further optimizing the outlier detection mechanisms and exploring their applicability to other datasets and use cases.
\\\
Additional illustrative information and experimental details are provided in the appendix.

\subsection*{Appendix}

\begin{figure}[H]
    \centering
    \begin{subfigure}[b]{0.8\textwidth}
        \centering
        \includegraphics[width=\textwidth]{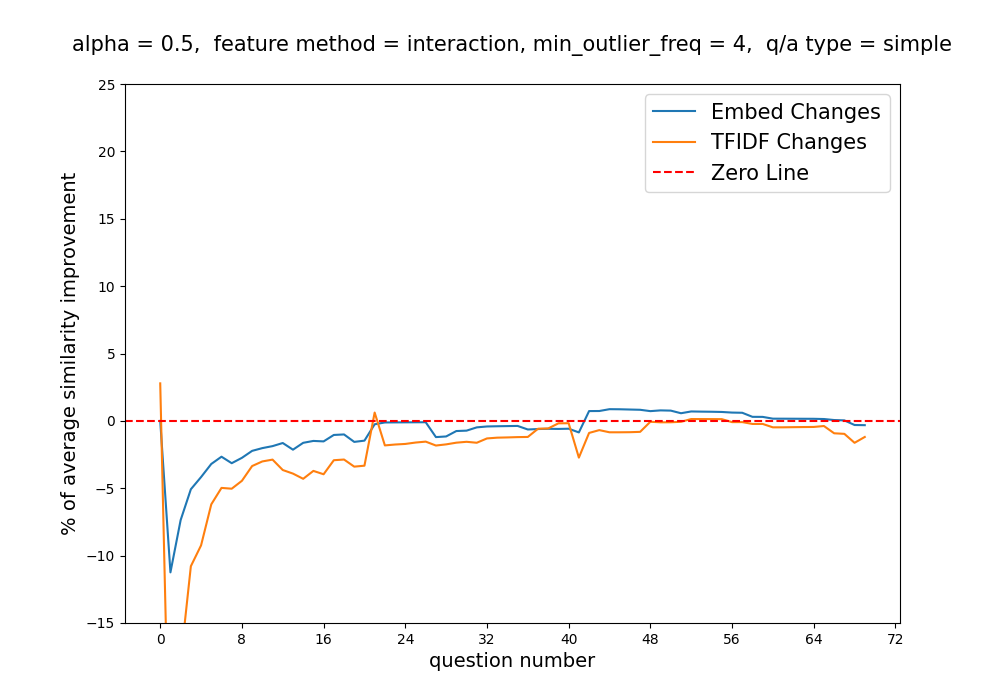}
        \label{fig:image1}
    \end{subfigure}

    \begin{subfigure}[b]{0.8\textwidth}
        \centering
        \includegraphics[width=\textwidth]{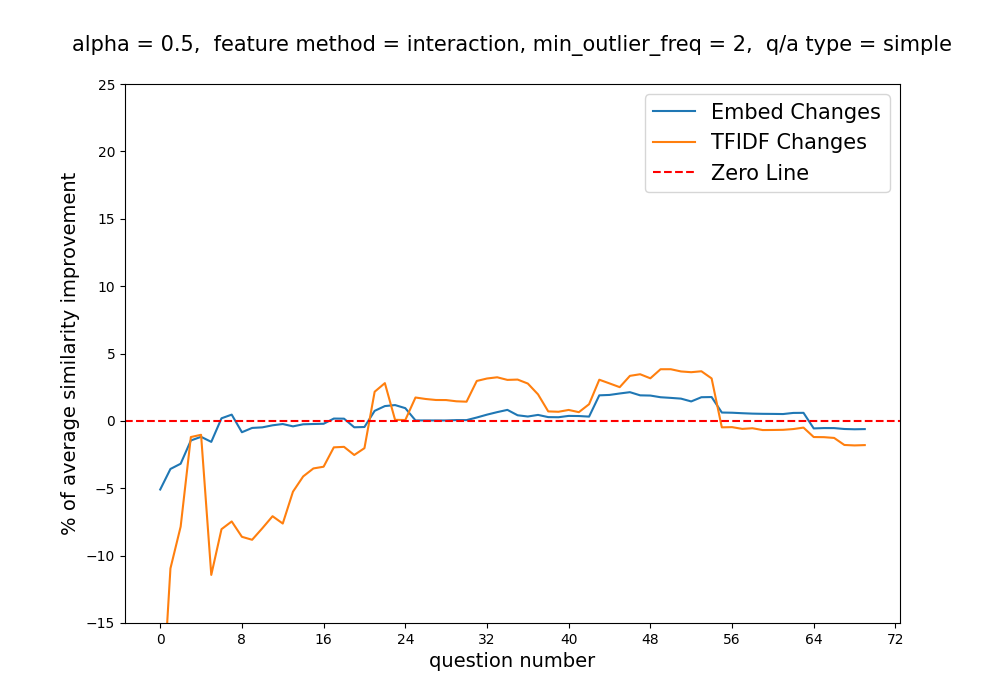}
        \label{fig:image2}
    \end{subfigure}

    \caption{The average values of similarity changes are presented as the number of processed questions increases (Part 1).}
\end{figure}

\begin{figure}[H]\ContinuedFloat
    \centering

    \begin{subfigure}[b]{0.8\textwidth}
        \centering
        \includegraphics[width=\textwidth]{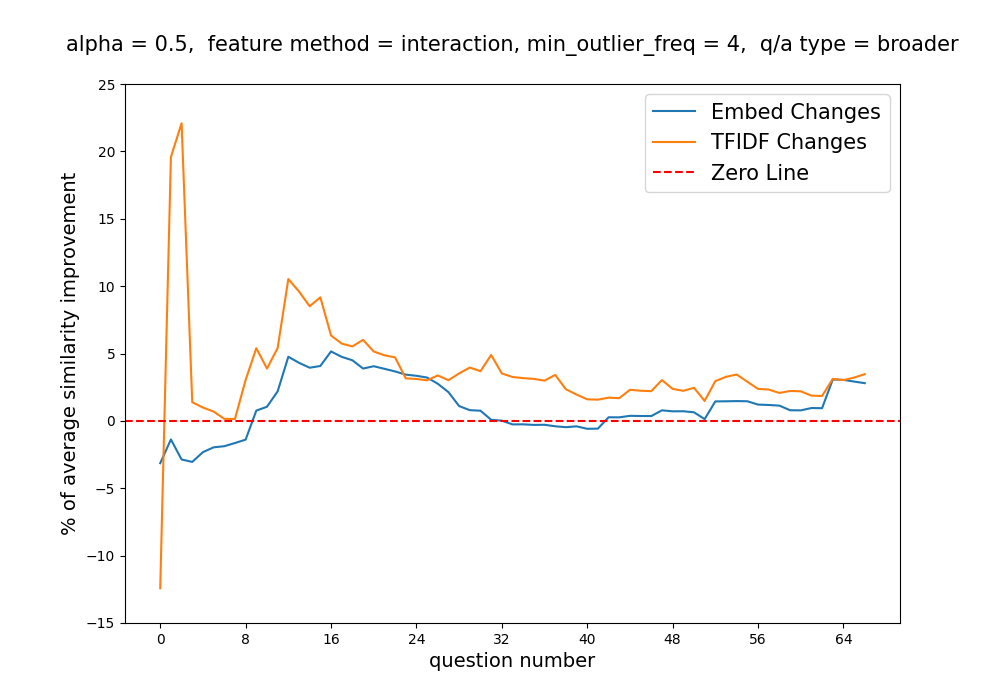}
        \label{fig:image3}
    \end{subfigure}

    \begin{subfigure}[b]{0.8\textwidth}
        \centering
        \includegraphics[width=\textwidth]{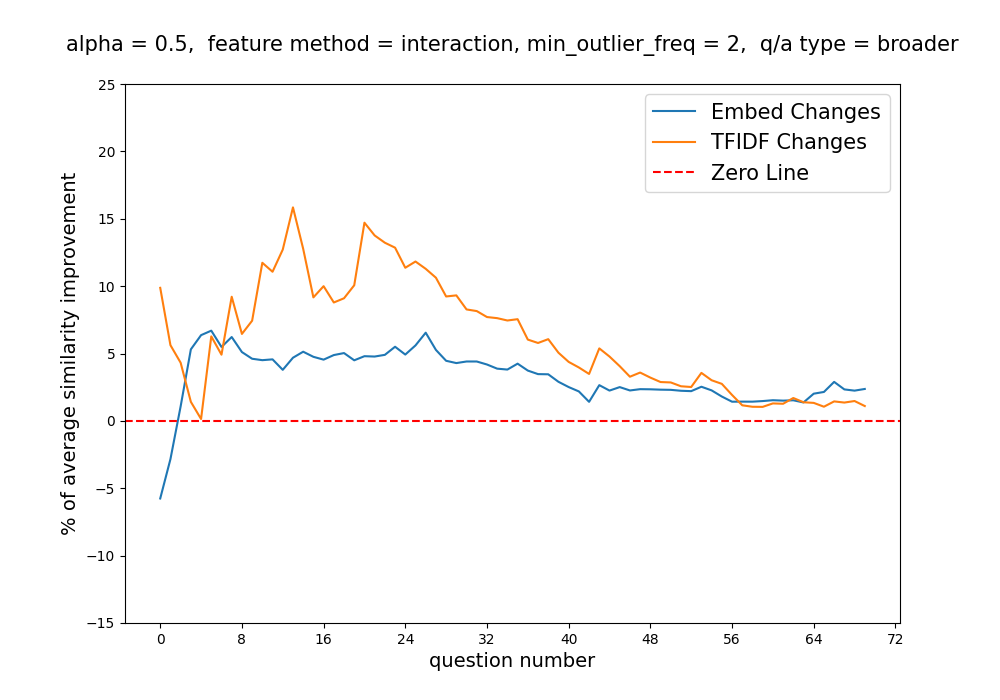}
        \label{fig:image4}
    \end{subfigure}

    \caption{The average values of similarity changes are presented as the number of processed questions increases (Part 2).}
\end{figure}

\begin{figure}[H]\ContinuedFloat
    \centering

    \begin{subfigure}[b]{0.8\textwidth}
        \centering
        \includegraphics[width=\textwidth]{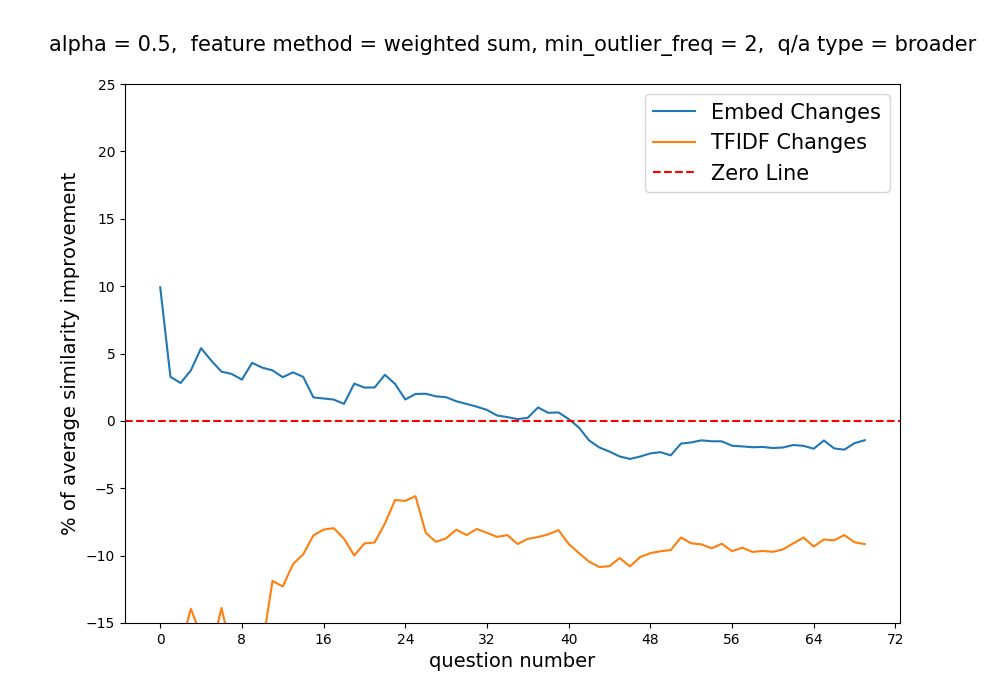}
        \label{fig:image5}
    \end{subfigure}

    \begin{subfigure}[b]{0.8\textwidth}
        \centering
        \includegraphics[width=\textwidth]{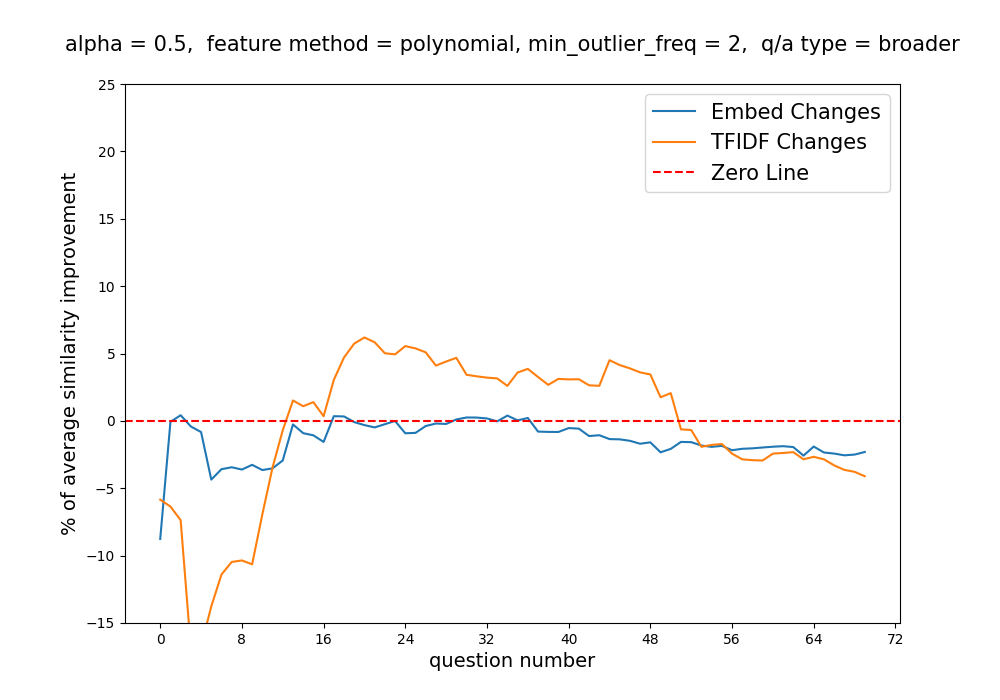}
        \label{fig:image6}
    \end{subfigure}

    \caption{The average values of similarity changes are presented as the number of processed questions increases (Part 3).}
\end{figure}

\begin{figure}[H]\ContinuedFloat
    \centering

    \begin{subfigure}[b]{0.8\textwidth}
        \centering
        \includegraphics[width=\textwidth]{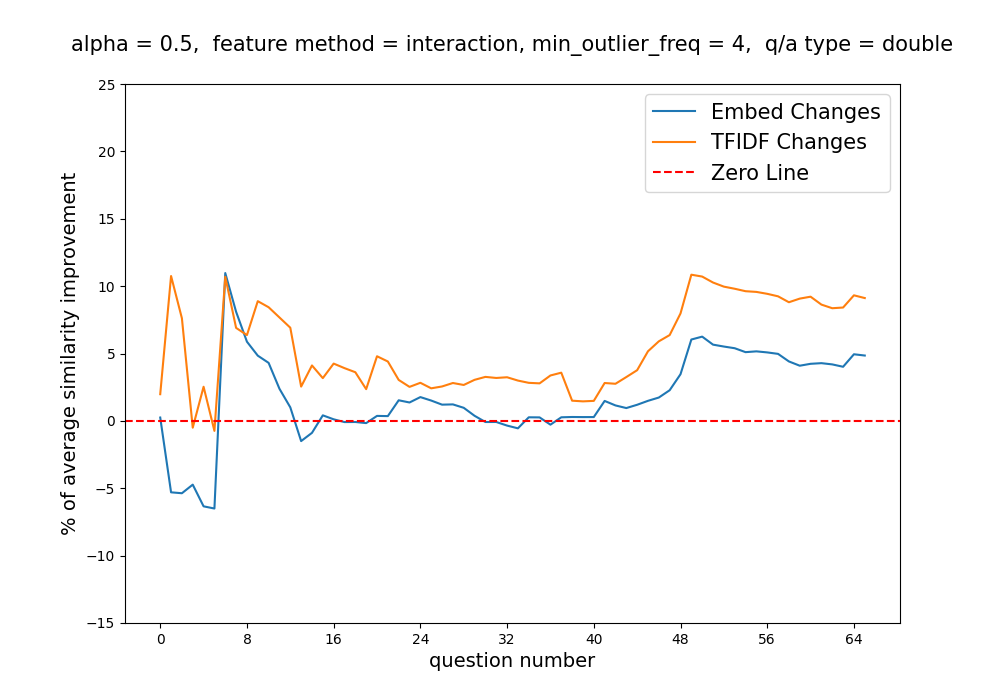}
        \label{fig:image7}
    \end{subfigure}

    \begin{subfigure}[b]{0.8\textwidth}
        \centering
        \includegraphics[width=\textwidth]{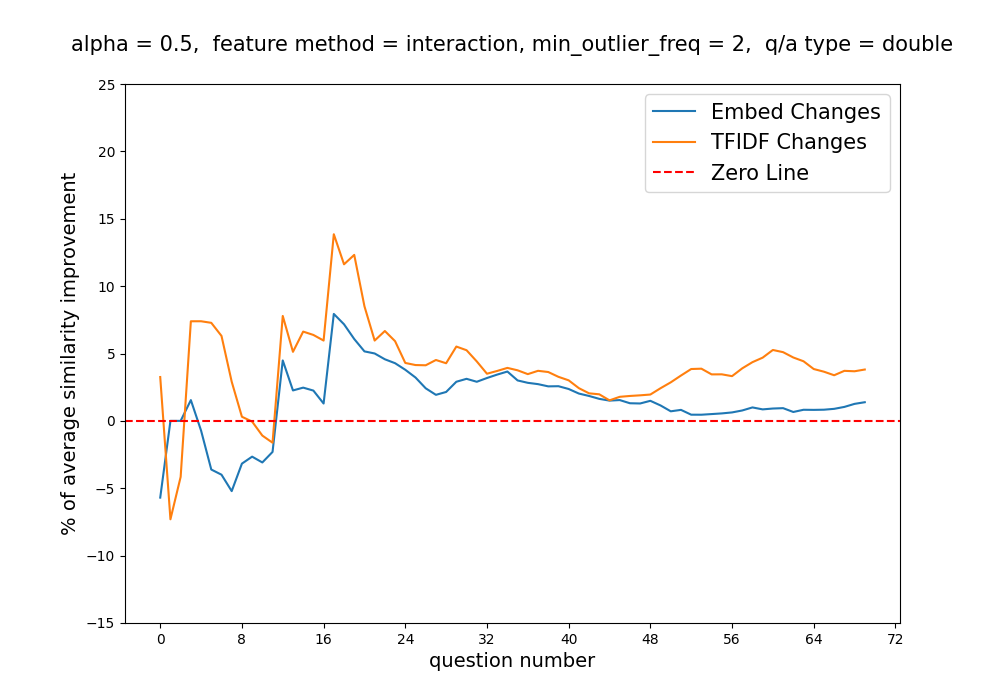}
        \label{fig:image8}
    \end{subfigure}

    \caption{The average values of similarity changes are presented as the number of processed questions increases (Part 4).}
\end{figure}

\begin{figure}[H]\ContinuedFloat
    \centering

    \begin{subfigure}[b]{0.8\textwidth}
        \centering
        \includegraphics[width=\textwidth]{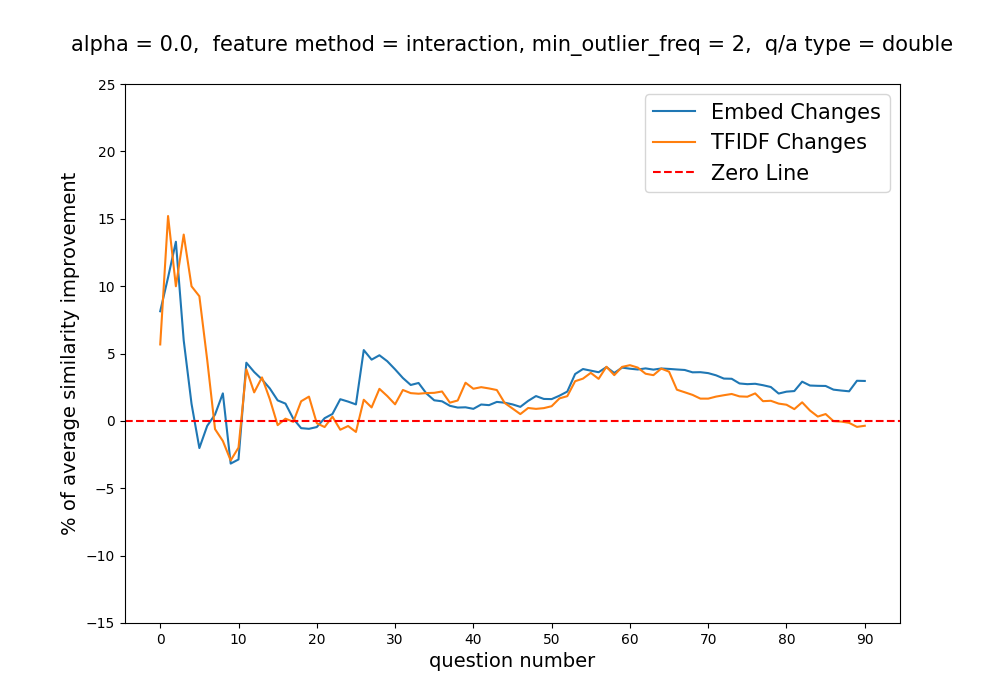}
        \label{fig:image9}
    \end{subfigure}

    \begin{subfigure}[b]{0.8\textwidth}
        \centering
        \includegraphics[width=\textwidth]{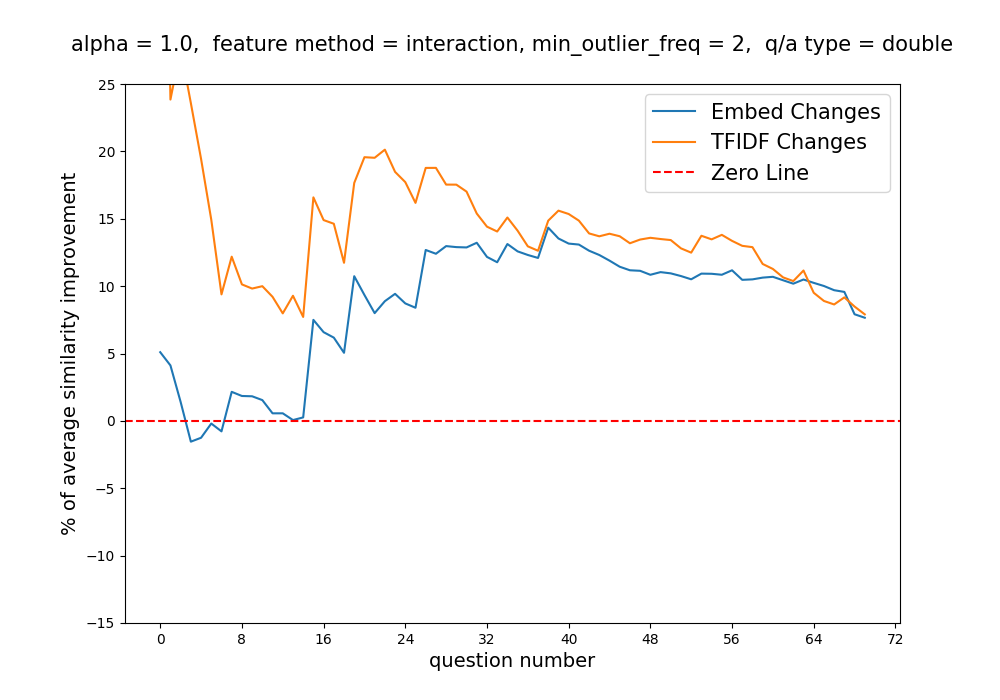}
        \label{fig:image10}
    \end{subfigure}

    \caption{The average values of similarity changes are presented as the number of processed questions increases (Part 5).}
\end{figure}

\end{document}